# A 481pJ/decision 3.4M decision/s Multifunctional Deep In-memory Inference Processor using Standard 6T SRAM Array

Mingu Kang, Sujan Gonugondla, Ameya Patil, and Naresh Shanbhag

Dept. Electrical and Computer Engineering, University of Illinois at Urbana-Champaign

**Abstract**

This paper describes a multi-functional deep in-memory processor for inference applications. Deep in-memory processing is achieved by embedding pitch-matched low-SNR analog processing into a standard 6T 16KB SRAM array in 65 nm CMOS. Four applications are demonstrated. The prototype achieves up to 5.6X (9.7X estimated for multi-bank scenario) energy savings with negligible (≤1%) accuracy degradation in all four applications as compared to the conventional architecture.

Emerging inference applications require processing of huge data volumes [1]. A conventional inference architecture (Fig. 1) implements memory access, data transfer from memory to processor, data aggregation, and slicing. In such architectures, memory access energy dominates, e.g., an 8-b SRAM read access and an 8-b MAC consumes 5pJ and 1pJ in 65nm CMOS, respectively. Additionally, the memory-processor interface presents a severe throughput bottleneck. Deep in-memory signal processing concept was proposed in [2] to overcome these challenges by embedding mixed-signal processing in the periphery of the SRAM bit-cell array (*BCA*). However, an IC implementation needs to address a host of new challenges including the stringent row & column pitch-matching requirements imposed by the *BCA* without altering its storage density or its read/write functionality, and enabling multiple functions with mixed signal circuitry. Recently [3], a single function, 5×1-b in-memory classifier IC has been demonstrated.

The proposed deep in-memory inference architecture has four stages (Fig.1): 1) multi-row functional read (*MR-FR*), 2) bit-line (*BL*) processing (*BLP*), 3) cross *BL* processing (*CBLP*), and 4) ADC and slicing. The *MR-FR* accesses multiple rows in one pre-charge cycle using pulse-width modulated word-line (*PWM-WL*) signals to generate a *BL* voltage drop proportional to a weighted sum of multiple bits stored in multiple rows in the column, and also performs word-level add/subtract. The *BLP* implements reconfigurable column pitch-matched mixed-signal circuits to execute computations such as multiply/absolute value/comparison on the *BL* voltages, in a massively column-parallel fashion. The *CBLP* aggregates the *BLP* outputs into a scalar which is sliced to obtain the final decision. The *BLP* and *CBLP* can be reconfigured to operate the architecture in either a dot product (*DP*) mode or Manhattan distance (*MD*) mode. Reconfigurable stages enable multiple functions (Fig. 1 table) including normal read/write.

The chip architecture (Fig. 2) includes a digital controller (CTRL) and a CORE. The normal read/write circuitry (lower) and in-memory processing blocks (upper) are physically separated to maintain functionality. Functional *WL* drivers generate *PWM-WL* signals while the reconfiguration word (*RCFG*) reconfigures local controllers in the CTRL. The in-memory processing chain is pipelined to enable *BL* pre-charge when the *MR-FR* step is complete. The architecture processes 128 8-b words per access cycle requiring two consecutive access cycles to process a 256-dimensional vector with 8-b elements. Thus, two consecutive *CBLP* outputs are sampled on different sampling capacitors and charge-shared before conversion by the ADC. Four 8-bit single-slope slow but energy efficient ADCs execute in parallel to process 36 128-dimensional vectors/µs.

The *MR-FR* step (Fig. 3) generates *BL* swing $\Delta V_{BL}$ proportional to binary-weighted bits ($d_i$) in a column [2] via the use of *PWM-WL* signals. An 8-b array data ($D$) and streamed input ($P$) precision is chosen to satisfy the requirements of many inference applications. The longest *PWM-WL* pulse width with $V_{WL} < V_{DD}$ is chosen to be less than 40% of *BL* RC time constant in order to ensure sufficient linearity and prevent destructive read [2]. The shortest pulse width needs to be <250ps while driving a large RC *WL*, which is challenging due to the row pitch-matching constraints. Hence, sub-ranged read is proposed where 4 MSBs and 4 LSBs are stored in adjacent columns (column pair), and read simultaneously on $BL_{MSB}$ and $BL_{LSB}$. Then, the charge on $BL_{MSB}$ is shared with 1/16 of $BL_{LSB}$ charge via switches $Ø_{con}$ and $Ø_{merge}$. Capacitors attached to *BL*s enable fine-tuning of the 1/16 capacitance ratio. The sub-ranged *MR-FR* (Fig. 3) achieves a maximum INL = 0.03 LSB.

In the *MD* mode, the *MR-FR* enables D-to-A conversion, and replica cell read performs word-level add/subtract by reading $P$ ($\bar{P}$ for subtract) from the replica bit-cell array simultaneously with $D$ (Fig. 3) [2]. The replica bit-cell array stores streamed data $P$ and can be written directly by *write BL* (*WBL*) reducing energy and latency overheads.

The *BLP* (Fig. 4) can be reconfigured to operate in either the *DP* or the *MD* mode for dot product or absolute computation, respectively. In the *MD* mode, an analog comparator and a mux is used to obtain the absolute value, and the multiplier circuit is reconfigured as a *BL*-wise sampler. In the *DP* mode, the comparator is bypassed and *BLB* is chosen by the mux. The mixed-signal capacitive multiplier [4] uses identical bit capacitors to meet the column pitch constraints necessitating sequential processing of multiplicand bits ($p_i$) and thereby limiting the throughput. Sub-ranged multiplication alleviates this problem by employing two 4-b MSB/LSB multipliers operating in parallel. The *BLP* outputs are charge-shared in the *CBLP* and sampled and later converted into a digital value by the ADC. In the *CBLP*, the MSB and LSB rails are charge-shared by first opening $Ø_{con\_rail}$ and then closing $Ø_{merge\_rail}$ to obtain the final output as a weighted sum. The measured accuracy of *BLP* and *CBLP* (including *MR-FR*) (Fig. 4) shows that the maximum error magnitude in the *DP* (*MD*) mode is 5.8% (8.6%) of output dynamic range.

The proposed architecture requires 16X fewer read accesses (and precharges) as compared to the conventional architecture for a fixed volume of data, resulting in up to 5.8X throughput enhancement. This is because *MR-FR* and *BLP* process data in massively parallel manner (128 8-b words per precharge) whereas the normal SRAM mode fetches only 8 8-b words through 4:1 column muxing. Smaller $\Delta V_{BL}$ and fewer read access reduces data access energy. Charge redistribution-based low-swing computation adds to the energy savings.

Fig. 5 (left) indicates that CORE energy and decision accuracy trade-off with $\Delta V_{BL}$. For binary (64-class) decisions, $\Delta V_{BL}$ > 15 mV (> 25mV) results in > 90% detection accuracy, and the CORE energy reduces by 0.2pJ (0.4pJ) per 20mV reduction in $\Delta V_{BL}$. Energy breakdown in Fig. 5 (right) indicates that much of the savings is due to *MR-FR*. The CTRL energy will be amortized in a multi-bank scenario. The measured energy in *DP* (*MD*) mode is 5.6× (3.7×) smaller than conventional architecture, with savings up to 9.7× (5.4×) in a multi-bank scenario.

The multifunctional IC (Fig. 6) implements four different algorithms with 2, 4, and 64-class decisions in a 512×256 SRAM array achieving better decision accuracy and comparable EDP (scaled for 65nm) than single function ICs [1,3]. The chip micrograph (Fig. 7) shows that the deep in-memory circuitry incurs an area overhead of 25% not counting the CTRL.

**Acknowledgement**

This work was supported by Systems on Nanoscale Information fabriCs (SONIC), one of the six SRC STARnet Centers, sponsored by SRC and DARPA.

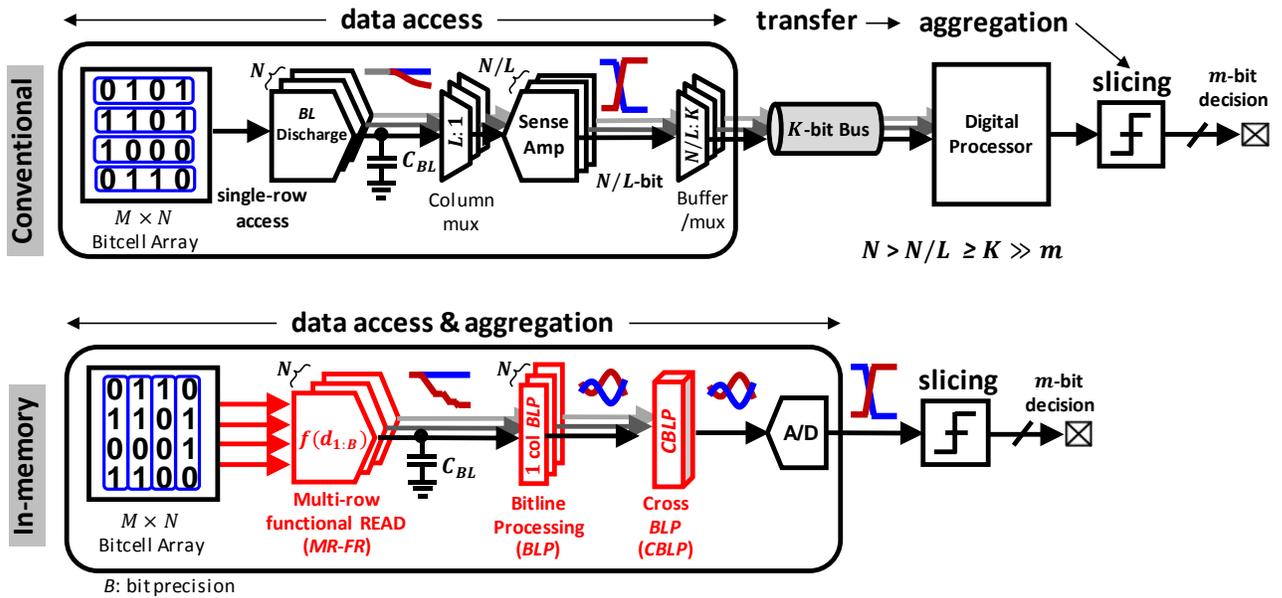

Figure 1: Conventional and proposed multi-functional deep in-memory architecture for inference applications.

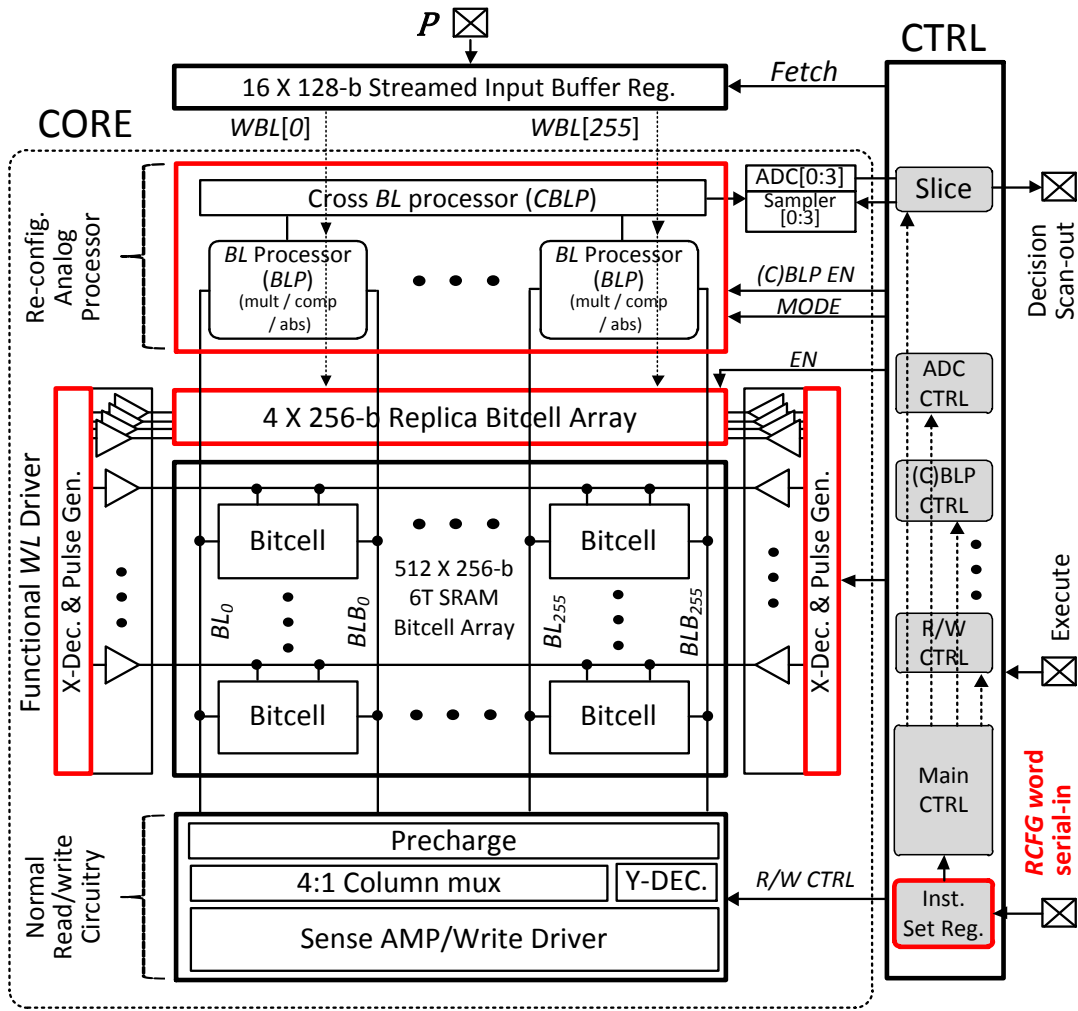

Figure 2: Deep in-memory processor architecture.

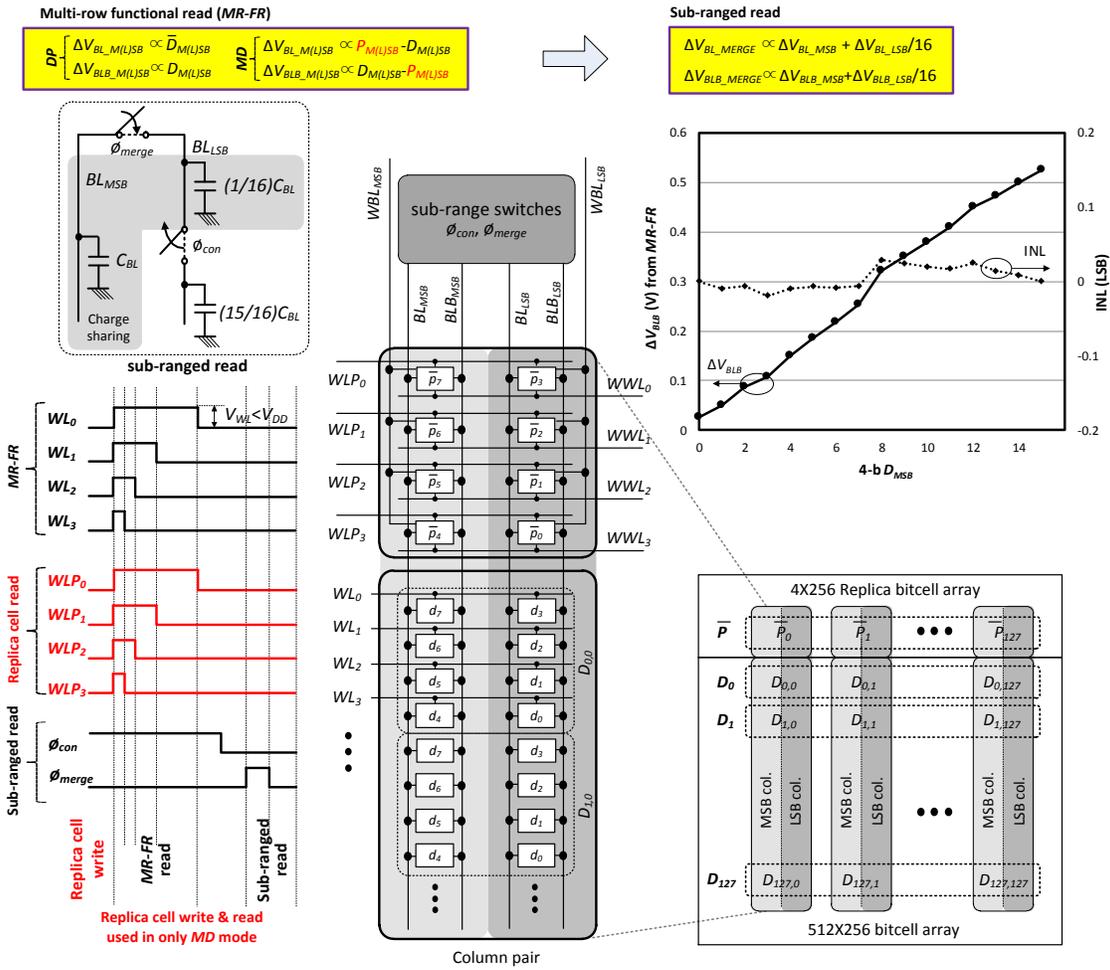

Figure 3: Sub-ranged multi-row functional read (*MR-FR*) and measured accuracy.

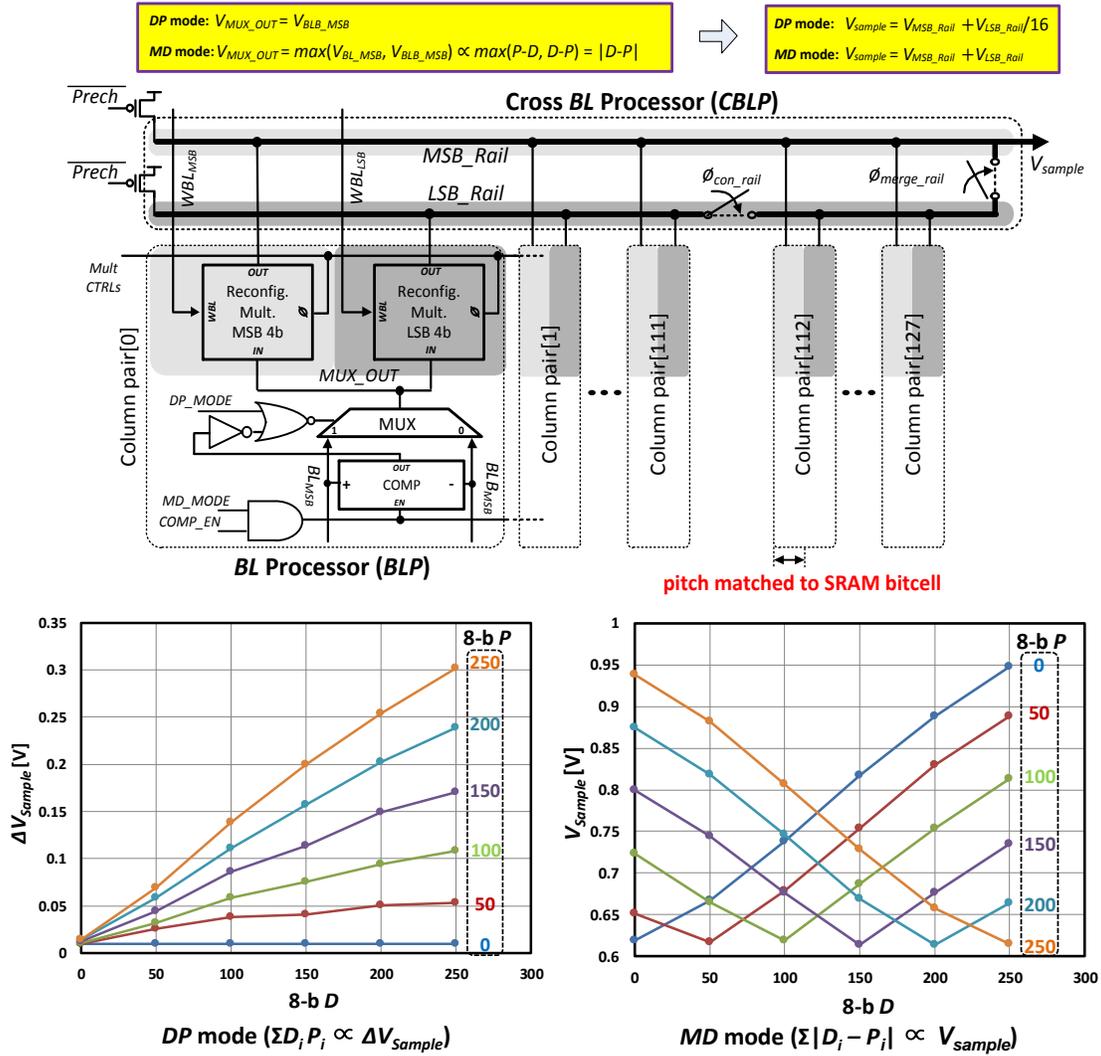

Figure 4: Pitch-matched *BL* processing (*BLP*), cross *BL* processing (*CBLP*), and measured accuracy (@ $D_0 = D_1 = \ldots = D_{255}$, $P_0 = P_1 = \ldots = P_{255}$).

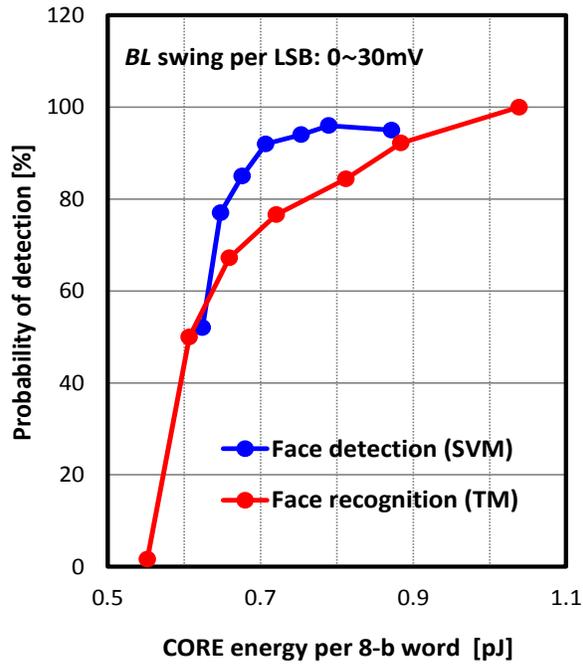 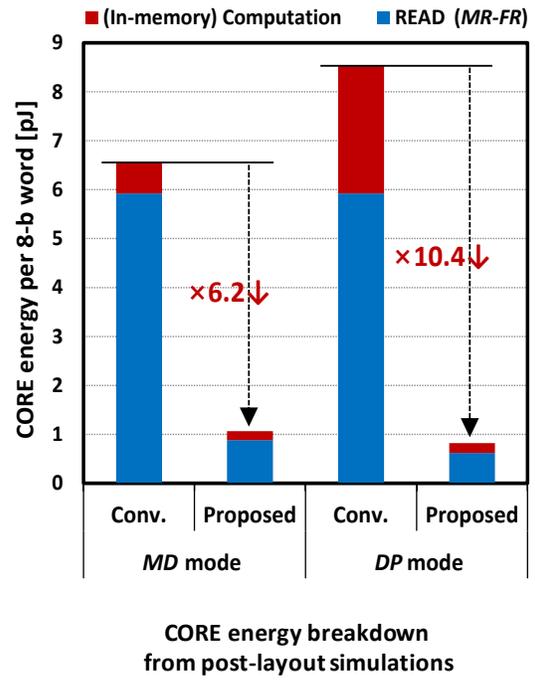

Figure 5: Measured energy vs. detection accuracy trade-offs, and CORE energy breakdown.

|   | Application | Algorithm | Dataset (*P*: Query input, *D*: data stored in array) |
|---|---|---|---|
| 1 | Face detection (binary class) | Support Vector Machine (*SVM*) | MIT CBCL dataset *P*: 23 X 22 8-b pixel image (face / non-face), 100 query inputs tested *D*: Feature extractor and classifier combined 23 X 22 8-b word coefficient |
| 2 | Event (Gun shot) detection (binary class) | Matched filter (*MF*) | *P*1: Gun shot sound contaminated by AWGN with 3 dB SNR or *P*2: Only AWGN with equal power of "signal + AWGN" in *P*1 total 100 query inputs tested *D*: gun shot mono sound data with 256 8-b words |
| 3 | Face recognition (64 classes) | Template matching with Manhattan distance (*TM*) | MIT CBCL dataset, 16 X 16 8-b pixel image for *P* and *D* *P*: one of the 64 candidate faces in *D*, 64 query inputs tested *D*: 64 candidate faces |
| 4 | Hand-written number recognition (4 classes) | K-nearest neighbor (*KNN*) | MNIST dataset, 4 classes from "0" to "3" (due to array size limit) 16 X 16 8-b pixel image for *P* and *D* *P*: image from 4 classes, 100 query inputs tested *D*: 16 images per class |

|   | Tech. (nm) | # of Algorithms | Memory size | Precision (b) | Decision Throughput (Decisions/s) | Decision Energy (pJ/decision) | Decision EDP (fJ·s) | Accuracy |
|---|---|---|---|---|---|---|---|---|
| This work | 65 CMOS | 4 (*SVM, MF, KNN, TM*) | SRAM 512 X 256-b | **8x8** | *SVM*: 9.3M | 963.1 / 462.4† | 0.1 / 0.05† | 95 % |
|   |   |   |   |   | *MF*: 18.5M | 481.5 / 231.2† | 0.03 / 0.01† | 100 % |
|   |   |   |   |   | *TM*: 312.5K | 33.6K / 17.5K† | 107.3 / 56.0† | 100 % |
|   |   |   |   |   | *KNN*: 312.5K | 33.6K / 17.5K† | 107.4 / 56.0† | 92 % |
| 8-b digital* | 65 CMOS | synthesized dedicated processor per algorithm | SRAM 512 X 256-b | **8x8** | *SVM*: 1.7M | 4.5K | 2.6 | 96 % |
|   |   |   |   |   | *MF*: 3.4M | 2.2K | 0.6 | 100 % |
|   |   |   |   |   | *TM*: 54.3K | 93.0K | 1715.3 | 100 % |
|   |   |   |   |   | *KNN*: 54.3K | 93.0K | 1715.3 | 90 % |
| [1]†† | 14 Tri-gate | 1 (*KNN*) | 128 byte | **8x8** | 21.5M | 3.4K | 0.2 | Not reported |
| [3]** | 130 CMOS | 1 (Adaboost) | SRAM 128 X 128-b | **5x1** | 50M | 633.4 | 0.01 | 90 % |

\* memory (digital) energy & delay measured from prototype IC (post-layout simulations); † assumes a 32 bank configuration;
†† single function with SRAM memory access cost not included; ** single function with 1b weight vector

Figure 6: Application level gains in energy efficiency, delay, accuracy, and comparison with prior arts.

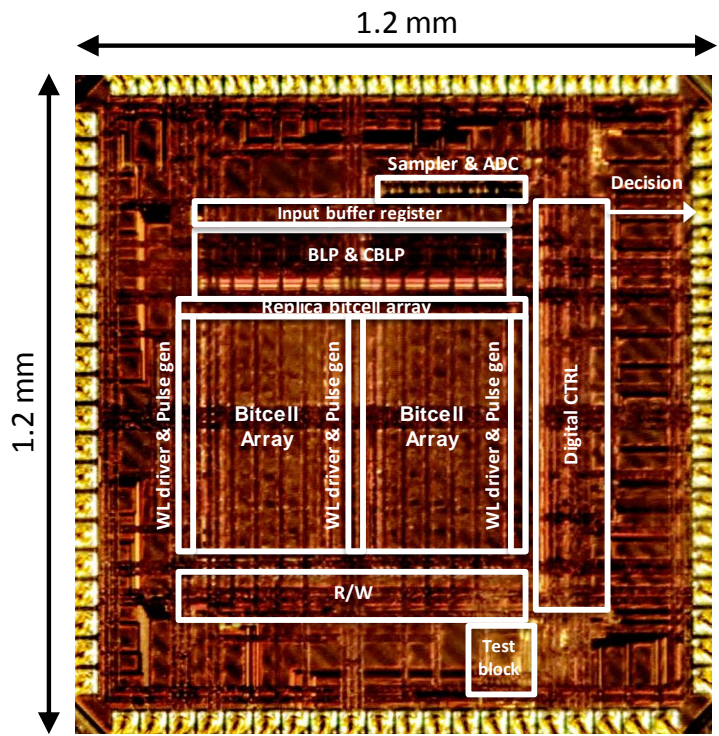

| Technology | 65 nm CMOS |
|---|---|
| Die size | 1.2 mm × 1.2 mm |
| CTRL operating freq. | 1 GHz |
| SRAM capacity | 16 KB (1 bank of 512 × 256-b) |
| Bitcell dimension | 2.11 × 0.92 um$^2$ |
| Supply voltage | CORE: 1.0 V, CTRL: 0.85 V |

| | | |
|---|---|---|
| Energy per decision (pJ) | *SVM* | 963.1 |
| | Matched filter | 481.5 |
| | *KNN* | 33.6K |
| | Template matching | 33.6K |
| Decision Throughput (Decisions/s) | *SVM* | 1.7M |
| | Matched filter | 3.4M |
| | *KNN* | 54.3K |
| | Template matching | 54.3K |

Figure 7: Die micrograph and chip summary.